 \documentclass{basi}
\begin{document}

\title[An X-ray view of quasars]{An X-ray view of quasars} 

\author[K.~P.~Singh]%
       {K.~P.~Singh$^1$\thanks{email: \texttt{Kulinderpal.Singh@tifr.res.in}}\\
       $^1$Tata  Institute of Fundamental Research, Homi Bhabha Road,
         Mumbai 400 005, India\\}

\pubyear{2013}
\volume{41}
\pagerange{\pageref{firstpage}--\pageref{lastpage}}

\date{Received 2013 April 30; accepted 2013 June 03}

\maketitle
\label{firstpage}
\begin{abstract}
I present an overview of observational studies of quasars of all types, with particular
emphasis on  X-ray observational studies.  The presentation is based on the most popularly
accepted unified picture of quasars - collectively referred to as AGN (active galactic nuclei) in this review.
Characteristics of X-ray spectra and X-ray variability obtained from various X-ray satellites over the
last 5 decades have been presented and discussed. The contribution of AGN in understanding the
cosmic X-ray background is discussed very briefly. Attempt has been made to provide up-to-date
information; however, this is a vast subject and this presentation is not intended to be 
comprehensive.
\end{abstract}
\section{Introduction}\label{i:Introduction}

Quasars discovered about 50 years ago by astronomers working with radio and optical telescopes
are the brightest sources of radiation in the sky covering a wide range of  wavelengths 
from radio - infra-red - optical - uv - X-rays to $\gamma$-rays. 
These cosmic objects appear as point like or ''quasi-stellar'' on photographic plates 
and are so powerful that they emit nearly 5 to 8 orders of magnitude more radiation than the normal galaxies. 
Their emission is concentrated in a very small, nuclear region of what are now known
as active galaxies. These also contain strong and broad optical emission lines.
The somewhat lower luminosity analogs of these objects were in fact discovered 
two decades earlier by Carl Seyfert and these are known as the Seyfert galaxies. 
The optical emission from them is characterised by peculiar, broad, high ionization emission lines 
and these are believed to originate from the nuclear regions of these galaxies, with the total light output 
(continuum + lines) exceeding the total light from the rest of the galaxies by 1 to 5 orders of magnitude. 
Such active galaxies make up about 1\% of all galaxies. Even lower luminosity active galaxies (albeit 
more luminous than the normal galaxies) exist and are dubbed as LINERS (Low
Ionisation Nuclear Emission Regions). LINERS are believed to be more numerous than the Seyfert types. 
All the different types of sources are collectively known as Active Galactic Nuclei and henceforth 
will generally be referred to as AGN, except while discussing some special properties.

Most of the AGN have very weak radio emission and are generally called Radio Quiet (RQ) quasars or RQ AGN
or sometimes QSOs (Quasi Stellar Objects).
About 10\% of the optically bright quasi-stellar objects are strong radio sources. These powerful radio sources
have been found to contain strong directed emissions, known as jets, and that is how these got the attention. 
It is this nature and their appearance in the optical that got them the name of "quasars" or Radio Loud (RL) AGN. 
There are several types of quasars known today depending on their observed characteristics in the optical.  
These are Blazars or BL Lac objects named after their highly variable prototype - BL Lac, 
FSRQs (Flat Spectrum Radio Quasars), HPQs (High Polarisation Quasars), and OVVs (Optically
Violent Variables). See Peterson(1997), Kembhavi \& Narlikar (1999), Krolik (1999) for 
a comprehensive discussion of different AGN types.

X-ray emission from AGN is ubiquitous, and was discovered in the early rocket flights in the mid-sixties,
just a few years after the discovery of quasars. High X-ray luminosity, comparable to that in the optical
was detected in almost all of them. Very high luminosity is subject to Eddington limit which
implies mass of greater than $\sim$ 10$^6$ L$_{44}$ solar masses, where L$_{44}$ is luminosity in unit
of 10$^{44}$ ergs s$^{-1}$. 
This combined with rapid variability, specially in X-rays, 
indicates that trillions of solar luminosities are generated in a region smaller than 10$^{13-15}$cm. 
The powerful source in AGN is thus extremely compact and much more efficient than stellar processes.
Accretion onto compact objects, which is known to be more efficient than nuclear fusion, is thus 
indicated. All these considerations suggest accretion onto a putative super massive black hole (SMBH)
of mass in the range of a million to a billion solar masses as the primary source of energetic
and powerful radiations from AGN. X-rays with their penetrative nature through large columns 
of gas and dust surrounding the SMBH are the most direct probe of the accretion processes 
responsible for this massive energy generation close to the SMBH. 

AGN dominate the X-ray sky. Nearly 80\% of all the X-ray sources in any medium-deep survey of 
the X-ray sky at high Galactic latitude are AGN, and are thus a major contributor to cosmic 
X-ray background that was largely unresolved until the early 1990s. 
These powerful AGN also have a significant impact on their environments and play a major role 
in process of galaxy formation via feedback through winds and ionisation of the interstellar media. 
New and unexpected populations are being discovered even now in the new and deeper X-ray surveys.  
In the local Universe, obscured AGN are seen to be more common than unobscured AGN by a factor of 
$\sim$3$-$4 (Risaliti et al. 1999). How does this ratio evolve at higher redshifts is an important 
question currently under study. 
These heavily obscured AGN are also very important for the understanding of the
cosmic X-ray background above a few keV, and are the targets of the studies 
of AGN in hard X-rays.
Apart from solving the mystery of the cosmic X-ray background 
the new surveys are suggesting the existence of more than one major
epoch of black hole mass accretion assembly in the history of the Universe.

In the following section (Section 2) I introduce the unified scheme of AGN structure (Fig. 1)
and characteristics of different types of AGN.  The unified scheme has been built 
up from multi-wavelength observations over the past few decades (Beckman \& Shrader 2012). 
In Section 3, I briefly describe the various X-ray missions
that have played, and are currently playing, a major role in the various milestones achieved in the
study of AGN.  Results obtained from X-ray spectral and timing measurements of AGN of various types
are described in detail in Section 4. The role of the AGN in explaining the cosmic X-ray background
is briefly described in Section 5, followed by a short discussion on the future directions in the
study of AGN.

\section {AGN structure: the unified scheme}

\begin{figure}
  \centerline{\includegraphics[angle=0,width=12cm]{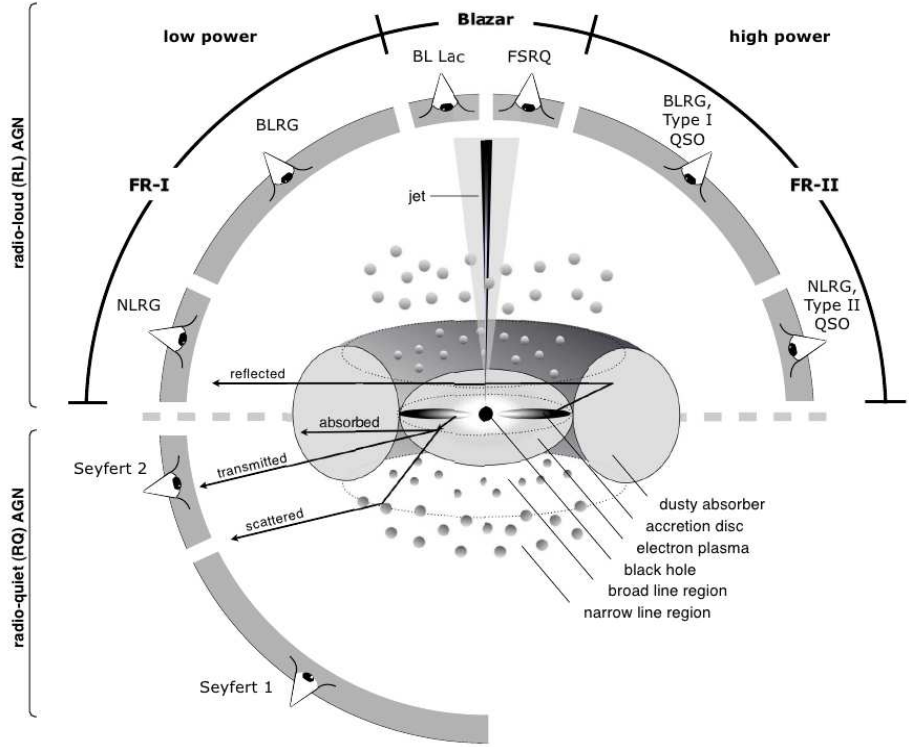}}
\caption{Schematic representation of our understanding of the AGN phenomenon in the unified scheme 
reproduced from Beckman \& Shrader (2012) (By permission of the authors). 
The type of object we see depends on the viewing angle, the presence or the absence of the jet,
and the power of the central engine in AGN. 
Note that radio loud objects are generally thought to display symmetric jet emission.} 
\end{figure}

The main physical components believed to exist in an AGN are displayed in the unified picture
shown schematically in Fig. 1.
The common powerhouse in each type of AGN is an accreting SMBH, 
while most of the differences amongst the various types can be ascribed to anisotropic AGN obscuring 
environments and their random orientations towards Earth (Fig. 1). The early ideas of unification
in which orientation and relativistic motion may play an important role in the observed properties
of AGN were suggested by Scheuer \& Readhead (1979) in their attempt to unify radio-loud and radio-quiet
quasars. These ideas were initially extended to lobe- and core-dominated radio loud quasars 
(Orr \& Browne 1982; Kapahi \& Saikia 1982) and later to radio-loud galaxies and quasars (Barthel 1989).

The idea of unification with an anisotropic AGN obscuring environment had its strongest evidence from 
spectro-polarimetric observations of the 
type 2 Seyfert NGC 1068 (Antonucci \& Miller 1985), in which broad permitted lines generated 
close to the nucleus and usually seen in type 1 Seyferts were discovered through polarised light. It was 
suggested that in type 2 Seyferts an optically-thick obscuring matter (mostly dust) blocks the line-of-sight 
to the nucleus of NGC 1068, and thus the broad permitted lines are not seen directly. A small 
fraction of the radiation can, however, get scattered into our direction if the matter surrounding the 
AGN has a torus geometry. The scattered light is polarised, thus suggesting that the reflector consists 
of electrons in a plasma, presumably ionised by direct AGN radiation. 
The narrow-line region in all AGN lies above the torus and is irradiated by the photo-ionising continuum.
Thus, the size and orientation angle of the torus to our line-of-sight is the key parameter that 
decides the appearance of an AGN. Broad permitted lines in type 2 AGN have also been discovered in the 
near-infrared by Goodrich et al. (1994), where the opacity is much lower than at optical wavelengths. 
The re-processed, warm, thermal radiation from the torus itself has been detected in the far-infrared 
(Wilman et al. 2000). It is the X-rays, however, that have the highest penetrating power through obscuring 
gas that is associated with the dusty torus surrounding the AGN.

Right at the centre of an AGN is an SMBH surrounding which is an accretion disk.
The size of the accretion disk around an SMBH of 10$^6$M$_{\odot}$ is of the order of light-days, which 
even at the distance of the nearest AGN, is too small to be resolved by the current generation of telescopes, 
since the resolution required is better than $\sim$1 milli-arcsecond. However, at radio frequencies it may
be possible to image maser emission from the disk with high angular resolution using Very Long Baseline
Interferometric techniques. For example, in the case of the AGN, 
NGC 4258, strong water maser emission has been used to map motions in the accretion 
disk (Miyoshi et al. 1995).  Accretion disks in AGN are usually optically-thick and physically-thin 
(Shakura \& Sunyaev 1973), and have temperatures in the range of $\sim$ 10$^5$ $-$ 10$^6$ K, making them 
sources of optical and ultraviolet radiation. This optical and UV continuum scatters off a corona of
very hot electrons close to and above the accretion disk. 
The corona of electrons is probably heated by processes similar to the heating of the Solar corona,
involving magnetic fields. Optical and UV photons from the disk undergo inverse-Compton scattering 
off hot electrons, and emerge as higher energy X-rays. Multiple scatterings within the corona 
boost the energy further, resulting in characteristic power-law non-thermal continuum extending 
from under 1 keV to several hundred keV. More details of the main features in AGN X-ray spectra are 
discussed in section 4.1.

The bulk of line emission from AGN, first used to classify the AGN, comes from discrete and patchy 
cloud-like structures much further out from the SMBH. These lines are generated mainly by 
photo-ionisation with some contribution from collisional excitation. 
The clouds closer to the SMBH have very broad permitted recombination lines 
(full width at half maximum: FWHM $\geq$3000 km s$^{-1}$), and also have blue broad-band 
optical continua. 
Thus, a dense (electron density of 10$^7$ $-$ 10$^{10}$ cm$^{-3}$) broad line region (BLR) is inferred to
exist on scales smaller than a parsec (pc) within the deep gravitational potential of SMBH, leading
to the large line-widths observed due to Doppler broadening. 
A technique called reverberation mapping that uses the delay time between continuum variation and
subsequent emission line intensity variations suggests the size of this region to be 
0.1L$_{46}^{0.5}$ pc (where L$_{46}$ is the photoionising luminosity in units of 
10$^{46}$ erg s$^{-1}$; Peterson 1993). 
The presence of BLR is responsible for designation of  type 1 for the AGN. 
AGN where the narrow (FWHM $\sim$ 500 km s$^{-1}$) permitted lines dominate are designated type 2 AGN 
and usually possess red continua. Intermediate types also exist which display both narrow and broad 
components to the permitted lines. All types show significant forbidden lines, but these are always narrow,
and thus suggesting the existence of a narrow line region (NLR), that extends over scales of tens to 
hundreds of pc, where the emitted lines are all narrow due to weaker gravity. 
The very existence of strong forbidden lines implies that densities in the NLR cannot exceed 
the critical density for collisional de-excitation, and are thus inferred to be low 
($\sim$ 10$^{4}$ cm$^{-3}$) from analysis of various nebular line species such as the 
[OIII] 4959, 5007 Angstroms doublet. Temperatures of the photoionised media are difficult to ascertain due 
to dependence of radiative cooling on metallicity, but values of $\sim$10$^{4}$ K are typical for the NLR; 
the BLR being hotter by a factor of $\sim$1.5 $-$ 2 (Osterbrock 1989).

In this unified scheme, quasars or Radio Loud (RL) AGN are objects with directed or relativistic jet emission
which produces strong radio emission via Synchrotron process. Blazars are interpreted as a 
subset of these radio-sources where the relativistic jet is aligned almost along 
the line of sight (Urry \& Padovani 1995). In other words, the class of blazars is represented by 
RL AGN observed very close to the direction of the relativistic jets ($<$ 10$^{o}$).

\section {X-ray missions and some milestones in the studies of AGN}
The first detections of X-rays from AGN were reported from 3C273 and M87 by Friedman \& Byram (1967) 
in a survey of the Virgo region carried out using a proportional counter on an Aerobee rocket. They also
suggested that combined contributions from such weak extragalactic sources could be responsible 
for the cosmic X-ray background already seen in the previous rocket flights. 
X-rays were subsequently detected from some Seyfert galaxies by the first dedicated X-ray satellite,
Uhuru, launched in 1970 (Gursky et al. 1971). NASA's High Energy Astronomy Observatory (HEAO-1) launched 
in 1977, carried scintillation detectors in addition to proportional counters, and provided 
complete surveys of the X-ray sky, especially towards high Galactic latitude regions (Piccinotti et al. 1982). 
Apart from identifying and studying X-ray emission from AGN (Dower et al. 1980; Mushotzky et al. 1980)
(and various other types of Galactic and extra-galactic objects) 
it also provided good spectral characterisation of the diffuse X-ray background 
radiation over the energy range 3 $-$ 50 keV (Marshall et al. 1980). 
A simple power-law was found to characterise the 3 to 50 keV X-ray emission of AGN. 
A soft X-ray ($<$ 1 keV) bump in the power-law X-ray spectrum of a Seyfert was also first discovered with the 
low-energy detectors on board HEAO-1 (Singh et al. 1985)
A veritable revolution in X-ray astronomy occurred when Einstein Observatory carrying the first X-ray 
telescope with focusing optics necessary to produce images (see Aschenbach (1985); Singh (2005, 2011) 
for an overview of X-ray imaging optics) was launched in 1978. Hundreds of extragalactic point-like X-ray 
sources, as well as diffuse and extended sources over 0.2 $-$ 4 keV were imaged with superb spatial resolution 
of a few arcsec. European Space Agency's EXOSAT and the Japanese Ginga satellites launched in the 
1980s carried out much longer observations of some AGN. EXOSAT confirmed the existence of soft bumps 
in the X-ray spectra (Turner \& Pounds 1989) and Ginga discovered the flattening of the hard 
X-ray spectra of AGN, indicating X-ray reflection (Pounds et al. 1990; Singh et al. 1990). 

Another major advance in X-ray astronomy came with ROSAT, or Roentgen satellite in the 
soft ($<$ 2 keV) energy range. In an all-sky survey, with its large collecting area 
and high resolution optics, ROSAT detected $>$100,000 X-ray sources (see Voges et al. 1999 and 
references therein).  It was able to completely resolve the soft X-ray background 
into discrete sources (mostly AGN) through ultra-deep observations (Lehmann et al. 2001).

ASCA (Tanaka et al. 1994) and BeppoSAX (Boella et al. 1997)
were important Japanese and Italian-led missions of the 
nineties respectively, that were responsible for detailed studies of AGN continua as well as 
Fe emission lines. ASCA with the first X-ray CCD camera in the focal plane of a light weight telescope
was the first to find evidence for broad emission line from Fe fluorescence in a deep observation 
of an AGN (Tanaka et al. 1995), suggesting the broadening of the red-wing to be due to extreme 
gravitational field very close to the SMBH.  X-ray flux variability at almost all time-scales from AGN 
was studied in detail by the dedicated satellite Rossi X-ray Timing Explorer (RXTE). 
RXTE, sensitive over the energy range 2 $-$ 200 keV, made several important discoveries 
(see McHardy 2010 and Section 4.2).

Chandra X-ray Observatory with its unprecedented  spatial resolution of $\sim$0.5 arcsec combined
with its high sensitivity over the 0.5 $-$ 8 keV range has led to the detection and identification of very
faint X-ray sources. One of the main results from the mission has been the resolution of the 
cosmic X-ray background radiation (CXB) up to approximately 8 keV (Mushtozky et al. 2000). 
It is now recognised that the bulk of the CXB is the summed emission from a distribution of 
redshifted, and predominantly-obscured AGN. 
Two sets of X-ray transmission gratings that are present in the optical path of Chandra have been used 
to observe large numbers of emission and absorption lines from X-ray sources, 
including AGN. For instance, the spectrum obtained for the AGN at the centre of NGC 3783 
shows hundreds of absorption features due to various ionic species 
of Fe, Mg, O, N, Ne, Si etc. (Kaspi et al. 2002).
XMM-Newton, ESA's major X-ray observatory, complements Chandra by having a much larger 
collecting area (a factor of $\sim$10 larger at 2 keV) over 0.3 $-$ 10 keV, but worse 
spatial resolution by a factor $\sim$10. Due to its large collecting area and good spectral resolution, 
XMM-Newton is able to search for spectral features in distant cosmic sources fainter than those 
detectable by any previous mission: e.g., Fe emission lines in distant AGN.

The latest to join the array of X-ray Observatories is the Nuclear Spectroscopic Telescope Array (NuSTAR), 
launched in 2012 (Harrison et al. 2013).  In the first result published using its ability 
to measure the high-energy (3 $-$ 80 keV) X-ray spectrum of NGC 1365 with unprecedented accuracy, 
Risaliti et al. (2013) report a clearer measurement of the broadening of 
the Fe fluorescence line that can be best explained as due to the spin 
($\geq$84\% of the maximum allowed value) of the SMBH in NGC 1365.  

\section {X-ray emission}
\subsection {X-ray spectra of radio-quiet AGN}
The many prominent features seen in the X-ray spectra of radio-quiet type 1 AGN  are shown in Figure 2 and
discussed below:

\begin{figure}
\centerline{\includegraphics[angle=0,width=12cm]{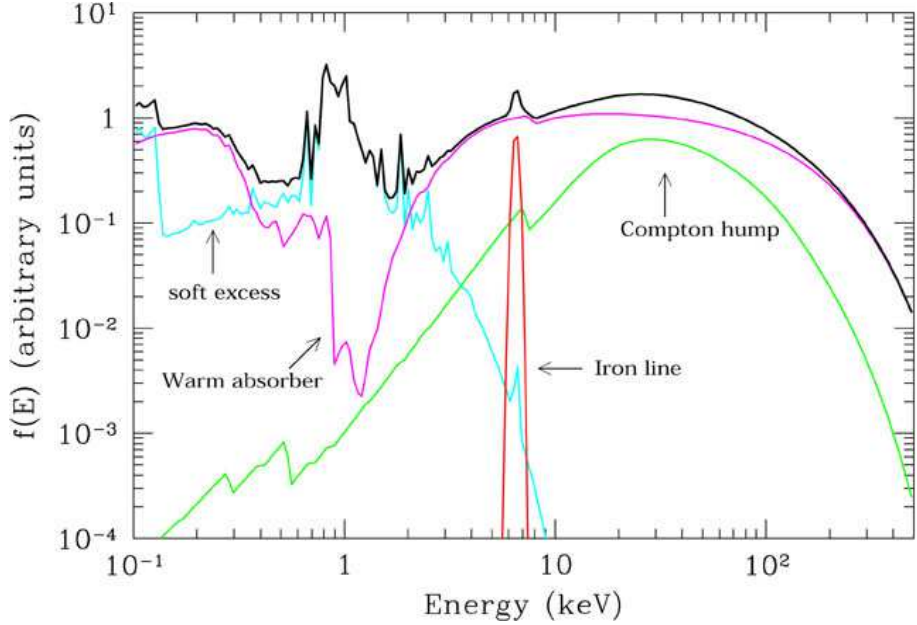}}
\caption{Average X-ray spectrum (thick black line) and its main components (thin grey lines)
 in an AGN of Type I reproduced from Risaliti \& Elvis (2004) (By permission of the authors). 
 "The primary continuum component is a power law with a high energy 
cut-off at E $\sim$100 $-$ 300 keV and absorption in soft X-rays by warm gas with 
N$_H$ $\sim$ 10$^{21}$ $-$ 10$^{23}$ cm$^{-2}$. A reflection component by cold matter is also shown. 
A narrow component due to iron K emission line at 6.4 keV can also be seen.  A ``soft X-ray excess'' component 
shown is believed to be thermal emission from a Compton thin plasma with temperature kT $\sim$ 0.1 $-$ 1 keV.}" 
\end{figure}

(i) Power-law: The primary component of an AGN spectrum is a
power-law in the 0.5$-$100 keV energy range. It is parametrized in terms of photon density,
N$_E$, proportional to E$^{-\Gamma}$ counts s$^{-1}$ cm$^{-2}$ keV$^{-1}$, where E is energy in keV and $\Gamma$ is the photon index. 
In the local Universe, an average value of the $\Gamma$ = 1.9 is observed (Nandra \& Pounds 1994), 
with typical range of 1.7 $-$ 2. The power-law is usually attributed to non-thermal processes arising due to 
multiple up-scatterings of accretion disk photons undergoing inverse-Compton interactions off 
thermal hot electrons in a corona above the accretion disk.

(ii) Soft Excess: An excess in soft X-rays above the power-law component component is observed in many (~50\%) of the
AGN type I, with a characteristic temperature in the range of kT $\sim$ 0.1 $-$ 1 keV 
(Singh et al. 1985; Turner \& Pounds 1989). This component  could be from the 
warm emitting gas located in the accretion disk, or in the broad line region (BLR) 
(perhaps the medium that confines the BLR clouds), or in a more distant region. 
This soft component could also be an extension of the big blue bump seen in the UV towards 
XUV and X-rays, caused by Compton scattering by hot electrons in the accretion disk corona 
(Czerny \& Elvis 1987).   AGN where the soft excess alone
dominates the X-ray emission, have been discovered and are known as the Narrow Line Seyfert Type 1 (NLS1)
galaxies since almost all of them have narrower permitted emission in the optical than seen in the 
regular Type 1 or broad line Seyferts.  
The power-law index of the X-ray emission from these objects is also found to be 
significantly steeper and inversely proportional to the width of the optical lines (Boller 1999). The steep power law
could be the result of domination of X-ray reflection (see below) in the innermost regions of the accretion flow 
(Fabian et al. 2002). NLS1 show extreme (large amplitude) and rapid (short time scale) X-ray variability 
among all the Type 1 AGN.

(iii) Hard X-ray Bump and Reflection: 
A broad bump, or excess luminosity, is seen in many AGN over $\sim$20 $-$ 40 keV above the 
canonical power-law. This has been attributed to reflection of photons 
(George \& Fabian 1991) briefly explained below: 
The hard power-law that is seen directly also irradiates the accretion disk, which produces backscattered
radiation, fluorescence, and recombination etc. This backscattered radiation or albedo
is known as reflection.
The reflection fraction depends on the ionisation state of the disk, but the overall effect is predicted 
to be a hardening or biasing of the spectrum towards higher energies, due to high albedo at 30 keV 
combined with photoelectric absorption of lower energy X-ray photons. A warm, ionized reflector usually present 
in $\sim$50\% of Seyfert 1 galaxies, is required in the central region of AGN for this to work.
The spectral shape of the reflected emission is the same as that of the incident spectrum. 
For a review of X-ray reflection in AGN, see Fabian \& Ross (2010) who also discuss how the reflection 
spectrum can be used to obtain the geometry of the accretion flow, particularly in the inner 
regions around SMBH.

(iv) Hard Cut-off: X-ray and $\gamma$-ray telescopes have discovered an exponential cut-off to the AGN 
power-law at energies between 100 $-$ 300 keV. At these high energies (approaching the electron rest-mass 
energy of 512 keV), photons lose more energy than they gain in each Compton scattering because of 
electron recoil, resulting in a roll-over of the spectrum.

(v) Low Energy Absorption:  Soft X-ray ($<$10 keV) emission from AGN is usually absorbed or 
modified before it reaches the observer through photoelectric absorption.   This is seen particularly 
if the line-of-sight passes through the obscuring torus. The obscuring  
column density that separates the obscured and unobscured (Type 1 or Type 2 respectively)
AGN is typically 10$^{22}$ cm$^{-2}$. 
The absorbing column-density itself is parametrized in terms of Hydrogen atoms per cm$^2$ integrated along 
the line-of-sight, though heavy elements such as O, Mg, Si, S and Fe are responsible for most of the 
opacity above 1 keV.  Discrete photoelectric edges are observed at the ionisation energies for 
these heavy elements (Morrison \& McCammon 1983). The obscuration for an e-folding decrease of 
transmitted radiation at 10 keV in the source rest-frame is $\sim$1.5 $\times$ 10$^{24}$ cm$^{-2}$, and these sources 
are called Compton-thick. Very little direct radiation is seen from these sources.
Additionally, at very high column-densities, multiple scatterings, as modelled by the Klein-Nishina cross-section, 
lead to a Compton down-scattering over the full X-ray spectral regime, and very little direct flux emerges 
above column-densities larger than 10$^{25}$ cm$^{-2}$ (Wilman \& Fabian 1999).  The effect of varying
absorption columns on the X-ray spectra of AGN is shown in Fig. 3 reproduced from Wilman \& Fabian (1999).

\begin{figure}
\centerline{\includegraphics[angle=0,width=12cm]{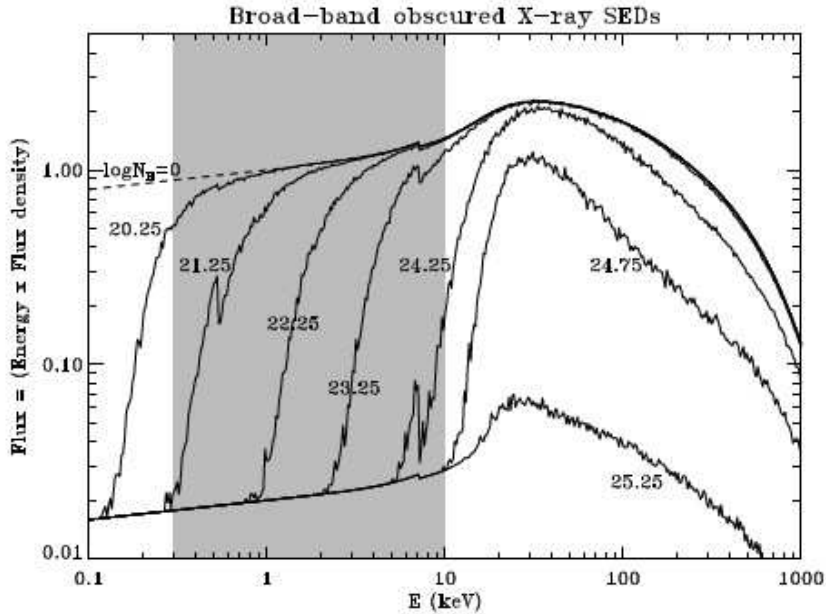}}

\caption{Rest-frame X-ray spectral energy distributions (SEDs) showing the effect of absorption  by material of solar 
abundance local to the source. Absorbers assumed have  column densities log(N$_H$) ranging from 20.25 to 25.25. 
Cross-sections are from Morrison \& McCammon (1983).  Compton down-scattering is modelled with 
the full Klein-Nishina cross-section.  The continuum spectra were modelled  with intrinsic AGN power-law with 
$\Gamma$=1.9 and include a reflection bump from cold material at solar abundance assuming 
uniform 2$\pi$ reflection and an inclination angle of 60$^\circ$, as well as an exponential cut-off at 360 keV.  
Two percent of the primary radiation is added to each SED to model the scattered component into the line-of-sight. 
See Wilman \& Fabian (1999) for details. The shaded region marks the X-ray spectral regime accessible 
to Chandra and XMM-Newton.  Courtesy: Wilman \& Fabian (1999).} 
\end{figure}

(vi) Iron Emission line: A prominent feature in the 2 $-$ 10 keV X-ray spectra of AGN is a narrow emission line
at 6.4 keV, arising from the Fe-K$_{\alpha}$ = 2 $-$ 1 transition of ``cold'' (i.e., Fe I $-$ FeXVII) iron. 
The EW of the line is $\simeq$100 $-$ 200 eV.
The line is believed to originate in the inner part of the accretion disk due to fluorescence.
The rotation of the accretion disk leads to the doppler-splitting of the line on the approaching and receding 
sides of the disk relative to the observer, i.e., the line photons lose energy due to strong gravitational redshift 
as the photons climb out of the deep potential well of the SMBH with additional distortions 
due to relativistic boosting of the blue (approaching) part of the spectrum relative to the red part
and due to the transverse doppler effect associated with time dilation. 
The resultant line profile has a blue peak and a bump with an extended red tail 
(see Fig. 4), each feature being sensitive to the characteristics of SMBH such as mass, spin etc.
Such asymmetric profiles have been calculated for lines emitted at a few Schwarzschild radii from non-rotating 
(George \& Fabian 1991; Fabian \& Ross 2010) and rotating (Laor 1991) SMBH.  
Evidence for a broad ``red wing'' extending to lower energies was first detected with ASCA (Tanaka et al. 1995).
However, the broad red wings have been seen with clarity only in a few objects 
e.g., MCG-6-30-15 (Vaughan \& Fabian 2004), 1H0707-495 (Fabian et al. 2009; Fig. 4). 
Most recently, broad-band X-ray (3 - 79 keV) observations of a low luminosity Seyfert NGC 1365 
with NuStar have also revealed these asymmetrically broadened Fe-line emission features and an 
associated Compton scattering excess at 10$-$30 keV (Risaliti et al. 2013). Using temporal 
and spectral analyses, Risaliti et al. disentangled the continuum changes due to time-variable absorption 
from reflection, which appears to arise from a region within 2.5 gravitational radii of a rapidly spinning 
black hole in NGC 1365.

An additional narrower component of the Fe-K line is also present in most AGN. The width of this ``narrow'' line 
(unresolved with ASCA, Chandra ACIS, or XMM-Newton MOS/EPIC CCDs) is 
$\leq$ 2000 km s$^{-1}$ when measured with the Chandra's high energy transmission grating (HETG). 
Similar to the width of broad emission lines in optical and UV, this component does not seem to vary 
with the continuum, even for delay times of days. 
The lack of variation and narrow width suggest an origin well beyond a few
Schwarzchild radii, although a smaller radius is also possible (Fabian et al. 2002). 
For a highly ionised reflector, the peak of the line can be shifted towards higher energies 
(6.7 keV or 6.96 keV for helium-like or hydrogen-like iron, respectively). 
Presence of both the narrow components, one from by a cold reflector and the other coming from an 
ionised reflector, is also possible. 
It is not possible to separate these two components using the resolution of $\sim$ 120 $-$ 150 eV at 6 keV  
provided by the CCDs in the ASCA or XMM-Newton. The higher resolution Chandra HETG spectrograph has a poorer sensitivity 
due to its low effective area of $\sim$40 cm$^2$ at 6 keV.

\begin{figure}
\centerline{\includegraphics[angle=0,width=10cm]{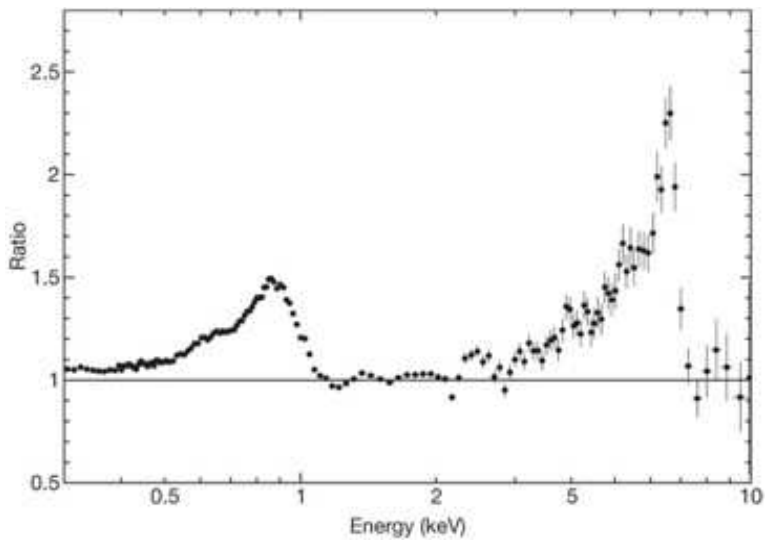}}

\caption{Ratio of the observed XMM-Newton spectrum of 1H0707-495 and a model consisting of a power law, a 
black body and two broad emission lines, where the normalizations of the broad lines have been set to zero 
to make the plot. Ionised Fe L and K emissions peaking in the rest frame around 0.9 keV and 6.5$-$6.7 keV 
with equivalent widths of 180 and 970 eV, respectively, can be seen. Error bars are 1$\sigma$. Courtesy: Fabian et al. (2009).} 
\end{figure}

The presence of both iron K and iron L emission in the spectrum of a narrow-line Seyfert 1 galaxy, viz., 
1H0707-495 has been reported by Fabian et al. (2009) and is shown in Fig. 4. 
Fabian et al. show that the presence of broadened and 
skewed iron L line is consistent with the prediction of disk reflection model, and that the relative strengths 
of the observed iron L and K lines agree well with predictions based on atomic physics with the iron abundance
being nine times solar. 

(vii) Absorption Lines: The availability of spectrographs with gratings onboard Chandra and XMM-Newton, 
has shown the presence of absorption lines due to a warm absorber component, which is probably 
formed by an outflowing gas. A high resolution spectrum obtained from a long observation of the 
Seyfert 1 galaxy NGC 3783 (Kaspi et al. 2002) is shown in Figure 5. 
Krongold et al. (2003) have shown that the observed lines can be produced by an absorber in two-phases 
which are in pressure equilibrium.

\begin{figure}
\centerline{\includegraphics[angle=0,width=14cm]{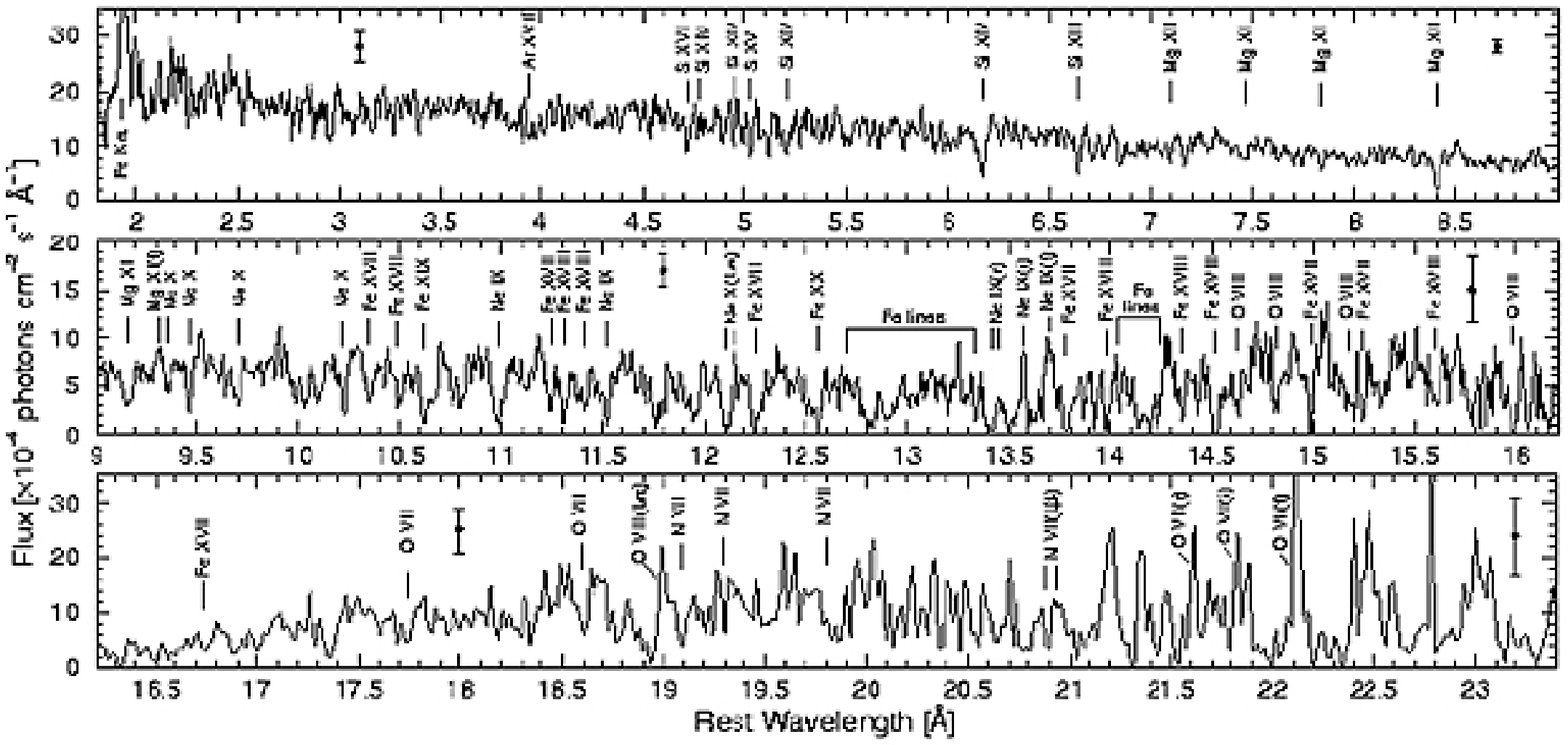}}

\caption{Combined MEG and HEG first-order 900 ksec spectrum of NGC 3783 obtained with Chandra,
 and binned to 0.01$\AA$.   The H-like and He-like lines of the identified ions are marked.  
The ions' lines are marked at their expected wavelengths in the rest frame of NGC 3783, and the blueshift 
of the absorption lines is noticeable. Courtesy Kaspi et al. 2002; Krongold et al. 2003.} 
\end{figure}

\subsection {X-ray variability of Seyfert galaxies}
X-ray emission from AGN is highly variable on a number of time scales: minutes, hours, days, months and years.
Since X-rays are produced from regions in close proximity to the SMBH, fast time scales of minutes to hours are
particularly useful in elucidating the location and size of the X-ray source in AGN.
The first detailed observations of AGN X-ray variability were made by EXOSAT (Lawrence et al. 1987). However,  
it was with RXTE observations in the last $\sim$16 years that a detailed picture has emerged and the reader is 
referred to McHardy (2010) where the author also compares the AGN with black-hole binary systems (BHBs) in the Galaxy.
On long time scales of months to years, BHBs are found to be in the ÔhardÕ, or ÔsoftÕ or ÔVHSÕ (Very High State) states.
In the ÔhardÕ state the medium energy X-ray flux is low and the spectrum is hard. 
In the ÔsoftÕ state, the medium energy X-ray flux is high and the spectrum soft, whereas in the VHS the spectrum 
is intermediate in hardness between the hard and soft states.  Type I AGN, on the other hand are usually 
found to have X-ray spectra that are similar to those of hard state BHBs (i.e. photon indices $\sim$ 2), 
and much harder than soft state BHBs. 

\subsubsection {Power spectral distributions and time lags between hard and soft X-rays}
The X-ray timing properties of the different spectral states are characteristically different and may
provide a better way to distinguish between different `states' than the spectral properties.
X-ray variability is usually quantified in terms of the power spectral density (PSD) of the X-ray light curve. 
The PSD of BHBs have a power-law spectrum P($\nu$) $\propto$ $\nu^{-\alpha}$
with a bend at high frequencies (slope, $\alpha$, goes from 2 to 1) in 
all their spectral states. In the hard state, the PSD shows an additional bend at low frequencies 
($\alpha$ going from 2 to 0). The frequency at which the bends occur refer to characteristic time scale of the systems
defined by the mass of the black hole and their accretion rate or ratio of luminosity to the the Eddington luminosity.
Measuring the PSDs of AGN are, however, difficult because it requires encompassing a wide enough frequency  
range to detect the equivalent of the bends seen in PSDs of BHBs and the data are irregularly sampled.
Assuming to first order that system sizes, and hence most relevant timescales, will scale with black hole mass, 
M$_{BH}$ , then as the break frequency $\nu_{B}$ is $\sim$10Hz for the BHB Cyg X-1 in the high state 
[M$_{BH}$ $\sim$ 10$-$20M$_{\odot}$],  the expected $\nu_{B}$ would $\sim$ 10$^{-7}$ Hz for an AGN 
with M$_{BH}$ =10$^8$ M$_{\odot}$. To measure such a bend requires a light curve stretching over 
a number of years. Uttley et al. (2002) describe a method that is used to derive PSDs from irregularly sampled 
light curves. Using this method for several AGN detects only one bend, from $\alpha$=2 at high frequencies to 
$\alpha$=1 at lower frequencies thus suggesting that only a soft state PSD is present in AGN (McHardy 2010)
and is consistent with AGN being very bright due to moderate to high accretion rates.  A couple of AGN, like Akn 564
and Ton S 180, appear to be accreting at nearly the Eddington rate and believed to be always in the VHS.
The PSD of these objects also show double bends but these are better described as two Lorentzians 
(McHardy et al. 2007; McHardy 2010).

Study of time lags between soft and hard X-ray light curves is a useful diagnostic to understand
the physical processes in AGN. Vaughan \& Nowak (1997) and Nowak et al. (1999) 
describe a statistic, the coherence function, 
derivable from the cross spectrum used to compute phase or time lags between two time series e.g., 
soft and hard X-ray light curves, as a function of frequency. 
Applying this method to Akn 564 and Ton S 180 showed that the nature of the lag changes with frequency
and this frequency is related to the transition point from one Lorentzian to another (McHardy 2010). 
In Akn 564, a positive lag  of $\sim$600s of hard X-rays for frequencies lower that 0.0004Hz changes 
to a negative lag of -11$\pm$4.3s at higher frequencies i.e.,  at high frequencies the soft X-rays lag the hard X-rays 
(McHardy et al. 2007; McHardy 2010; De Marco et al. 2013 and references therein). 
The hard lags are usually explained as due to multiple Compton scatterings of the low energy seed photons from the 
accretion disk to higher energies by a surrounding corona of very hot electrons. 
The soft lag would imply that in the very innermost part of the accretion disk corresponding to high frequencies in the
coherence spectrum, the soft X-rays originate mainly from reprocessing of harder X-rays by the accretion disk.

\begin{figure}
\centerline{\includegraphics[angle=0,width=14cm]{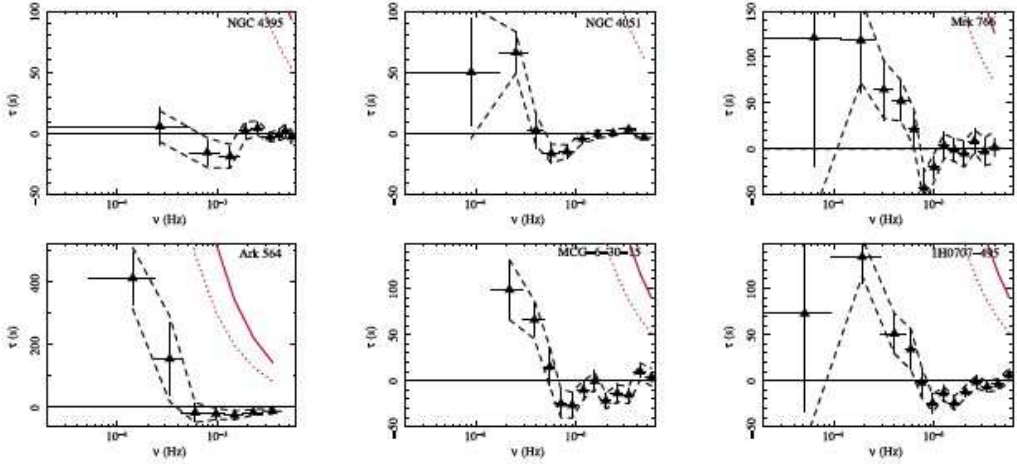}}
\caption{Hard and soft X-ray Lag$-$Frequency spectra of 6 the 32 sources reported by de Marco et al. (2013).
Detected soft/negative lag at $\>$2$\sigma$ confidence level are shown. The error bars are calculated as shown in 
Nowak et al. (1999). For further details see De Marco et al. (2013). Courtesy: de Marco et al. (2013)} 
\end{figure}

\subsubsection {Reverebration mapping and lags between reflected and direct X-rays}
The decomposition of the spectrum into direct and reflected emission components can be used in
interpreting the reverberation lag as a signature of the light travel time between the corona 
(the source of hard X-ray continuum) and the accretion disk (which reprocesses the flux into a reflection 
component with atomic features like the Fe-K$\alpha$ line).
Recently, a time delay of a few kilo-seconds was measured between the 2$-$3 keV band of the direct 
emission and the peak of the Fe K$\alpha$ line in NGC 4151 (Zoghbi et al. 2012). 
The change of lag with energy resembles the shape of an iron K$\alpha$ line broadened by relativistic effects.
In addition, the ÒlineÓ has a stronger blue horn at long timescales and a prominent red wing at short timescales, 
in a striking match to the expectation of a line emitted from a disk and distorted by relativistic effects 
(Fabian \& Ross 2010). 
Detection of Fe K lags has also been seen in 1H0707-495, IRAS 13224-3809, MCG$-$5$-$23$-$16 
and NGC 7314 (Kara et al. 2013; Zoghbi et al. 2013).  In the last two objects, the 6$-$7 keV band, 
where the Fe K$\alpha$ line peaks, lags the bands at lower and higher energies with a time delay of $\sim$1000s. 
These lags are fully consistent with reverberation of the relativistically broadened iron K$\alpha$ line. 
The measured lags, their time scale, and spectral modelling indicate that most of the radiation is emitted 
at $\sim$5 and 24 gravitational radii for MCG$-$5$-$23$-$16 and NGC 7314, respectively (Zoghbi et al. 2013).

Observations of Seyferts at luminosities L$\geq$10$^{-2}$L$_{Edd}$, show the presence of a very 
compact X-ray emitting source generally known as the accretion disk corona (McHardy et al. 2006). 
This region is believed to exist on top of "a geometrically-thin, optically thick accretion disk,  and 
may extend radially inwards reaching close to the innermost stable circular orbit (ISCO)"
(McHardy et al. 2006; Reis \& Miller 2013).
At lower luminosities, the hard X-rays probably come from a much larger region.
"The scale-invariant nature of the physics of accretion flows onto BHs leads to the lag time scale 
depending primarily on the mass of the central BH via t $\sim$ GM/c$^3$ = r$_{g}$/c, where r$_g$ is the 
Schwarzchild radius for mass M of a SMBH" (Reis \& Miller, 2013).
Reis \& Miller (2013) have estimated the probability distribution for the distances between the coronae and accretion 
disks.  They have used the probability distribution of the lag times and masses and applied it to a sample
of 17 radio-quiet Seyferts having measurements of soft lags and having luminosities $\geq$10$^{-2}$L$_{Edd}$.
According to Reis \& Miller: "14 out of 17 sources in that sample are indeed very compact, with distances 
in the range of  2$-$3r$_{g}$ above the accretion disk." The distances are $\leq$ 20r$_g$ for all the sources
even after all possible systematic uncertainties are considered, thus favouring the "coronal models 
in which bulk of the hard X-ray emission comes from magnetic reconnection in the innermost disk emitting 
via processes taking place in the compact base of a central, relativistic jet" (Reis \& Miller).

\subsection {Broad Absorption Line quasars (BALQSOs)}
Some quasars show Broad Absorption Lines or BALs in their optical and UV spectra. The lines are almost 
saturated and blue-shifted with outflow velocities as large as 0.1c (Weymann \& Morris 1991). 
These troughs can be as wide as 2000 $-$ 20000 km s$^{-1}$ and are due to resonance absorption in the gas. 
About 10\% of QSOs exhibit BAL troughs, but this is usually attributed to 
an orientation effect. No broad emission lines are observed with comparable widths, 
implying that the covering factor of the BAL region must be $<$20\% (Hamann et al. 1993). 
This observation, together with the fraction of quasars showing BALs suggests that most or possibly 
all quasars contain BAL-type outflows. Indeed, counting weaker absorption systems, such as mini-BALs, 
narrow absorption lines, and associated absorptions, the fraction of AGN 
with outflows is at least 40\%-60\% (Dai et al. 2008). 
BALQSOs may thus provide a unique probe of conditions near the nucleus of most QSOs. 
In the orientation scenario, we expect that the intrinsic X-ray emission of BALQSOs should be similar 
to that of normal quasars. Therefore, an understanding of BAL outflows is required for an understanding 
of quasars as a whole.

X-ray spectroscopy of BALQSOs is important to study the properties and dynamics of absorbing outflows. 
This has, however, been difficult since BALQSOs are faint X-ray sources. 
One of the first attempts to study a BALQSO in X-rays with a wide field of view detector onboard EXOSAT and ASCA
indicated heavy absorption in PHL5200 (Singh et al. 1987; Mathur et al. 1995), but was confused 
with another very faint quasar very close to PHL5200.  Subsequently, ROSAT observations with a finer point spread function revealed that BALQSOs as a class are quiet in soft X-rays, and have unusually steep 
optical-to-X-ray slopes, $\alpha_{ox}$ $\geq$1.94 relative to non-BAL QSOs  $\alpha_{ox}$ $\sim$1.6). 
Polarization studies also imply high column density absorption along the line of sight 
(Goodrich 1997). It is now believed that the BAL material has an absorbing column at least an 
order of magnitude larger than that estimated from UV spectra alone.
ASCA observations (Gallagher et al. 1999) pushed up the column density lower limit by an order of magnitude 
above the ROSAT value. The absorbing columns in some cases are at least 5 $\times$ 10$^{23}$, within a 
factor of three of being Compton thick. Deep (100 ks) ASCA observations of an optically bright BALQSO 
ruled out the possibility of BALQSOs being intrinsically X-ray weak, instead 
strong absorption was clearly indicated (Mathur et al. 2000).
The sharp point-spread function of Chandra enabled easy detections of the high-ionization 
BALQSOs (Green et al. 2001). The low-ionization BALQSOs, however, were found to be significantly more 
X-ray weak, implying either much larger absorbing column densities or intrinsically different X-ray characteristics.
The XMM spectroscopy confirmed that high-ionization BALQSOs are highly absorbed in X-rays, but the nature of 
their continuum shape remained unresolved (Grupe et al. 2003). No other property of BALQSOs, 
including radio-loudness and polarization fraction, was found to correlate with X-ray brightness 
(Gallagher et al. 2006). Almost all X-ray data of BALQSOs are consistent with continuum spectral 
slopes similar to normal quasars, but steeper slopes were not ruled out in most cases. 
A steep slope is implied in the case of the z = 5.8 BALQSO SDSS1044-0125 (Mathur 2001). 
All authors, however, agree on one important result: that strong X-ray absorption is a defining 
characteristic of BALQSOs.

BALQSOs were thought to be only radio quiet; however, radio-loud BALQSOs have been discovered 
in sensitive radio surveys (Becker et al. 1997). The fraction of BALs decreases with increasing 
radio luminosity (Shankar et al. 2008), partially explaining the initial failures when searching for 
BALs in radio-loud quasars.
Schaefer et al. (2004) found that XMM-Newton spectrum of radio-loud BALQSOs showed it to be brighter than
the radio-quiet BAL quasars and having a partial covering X-ray absorber. 
This implies that even radio-loud BALQSOs exhibit some soft-X-ray absorption as do radio-quiet BALQSOs. 
Miller et al. (2009), based on Chandra observations of BALQSOs, (note that these studies are based on X-ray detections with counts too few to perform spectroscopy) find that radio-loud BALQSOs are X-ray weak relative to 
their non-BAL counterparts.  Miller et al. also note that while some radio-loud BALQSOs have harder 
X-ray spectra than typical radio-loud quasars, some do not. 

\subsection {Radio loud AGN: quasars and blazars}
Radio-loud AGN: quasars and blazars, are very strong emitters of radiation at almost all the wavelengths: 
radio, mm, infra-red, optical, ultra-violet, X-ray and $\gamma$-rays from MeV to TeV. 
The Spectral Energy Distribution (SED) of blazars, usually shows two broad-band humps which are usually
believed to be produced via synchrotron (peaking in the infra-red to soft X-ray energy band) and 
Inverse Compton (IC) emission, dominating the hard X-ray to gamma-ray regimes, respectively. 
The lower energy (radio-to soft X-rays) emission component is synchrotron radiation from the jet 
whereas the high-energy component is believed to arise through the inverse-Compton scattering of soft photons 
by highly relativistic electrons in the jet plasma. These soft photons could originate from the local 
synchrotron radiation within the jet producing the Synchrotron Self-Compton (SSC) component
or they could originate from the nuclear IR/optical/UV emission producing the External Compton (EC) component 
(see Maraschi, Ghisellini \& Celotti 1992; Dermer \& Schlickeiser 1993; Sikora, Begelman \& Rees 1994).

The X-ray spectrum of a RL AGN is generally described by a power-law of the type:
N(E) $\propto$ E$^{-\Gamma}$.  Here, N(E) is the number of photons in 
units of s$^{-1}$ cm$^{-2}$ keV$^{-1}$, E is the energy, and $\Gamma$ is the photon index.  
However, the X-ray spectral shapes of blazars deviate from this simple power-law. 
These sources appear to have a curvature that shows a flattening in their soft X-ray spectrum 
(or a steepening in harder X-rays). 
The flattening in the very soft X-rays, with respect to a simple power-law of the primary spectrum,
has been interpreted by Page et al. (2005), Galbiati et al. (2005), and Yuan et al. (2006).  According to them it could be
either due to an excess absorption over the Galactic component (e.g. Cappi et al. 1997; Yuan et al. 2006), or due to 
a break in the intrinsic continuum (Sikora, Begelman \& Rees 1994; Tavecchio et al. 2007). 
It has been proposed that the absorption excess could be due a dependence of the hydrogen column 
density of the absorber (N$_{H}$ ) with the redshift z.  A correlation between N$_{H}$ and z was indeed reported 
by Yuan et al. (2006) from a sample of 32 RL sources observed with XMM-Newton.  
This correlation has been confirmed by Gianni et al. (2011) using a sample that is larger than the previous samples, 
and is based on analysis of broad-band (0.2 $-$ 100 keV) X-ray data obtained from Suzaku, Swift and INTEGRAL.
Gianni et al. report that the absorption at z$>$2 is higher when compared to the values reported previously.  
Extending the data to hard X-rays enabled them to detect even the highly absorbed sources 
(with N$_{H}$ $\geq$10$^{23}$ cm$^{-2}$ in the rest-frame of the source).  
The $\Gamma$ index for these sources had a maximum at a harder value ($\sim$ 1.4) 
with respect to the values ($\sim$ 2) obtained using the XMM data alone  (Gianni et al.).

In contrast with the radio quiet (RQ) AGN, the presence of reflection features - particularly the 
Fe K$_{\alpha}$ line and the Compton reflection bump at $\sim$ 20 $-$ 30 keV
are not always seen in RL AGN.  Both the iron line and the reflection bump can be 
absent or weak. Some of the RL AGN, known as Broad Line Radio Galaxies, however seem to show weak 
hard X-ray reflection features in observations with ASCA, RXTE and Beppo SAX 
(e.g. Sambruna, Eracleous \& Mushotzky 1999; Eracleous, Sambruna \& Mushotzky 2000; Grandi et al. 2001; 
Ballantyne, Ross \& Fabian 2002; Grandi \& Palumbo 2004; Grandi, Malaguti \& Fiocchi 2006).

\subsubsection {X-ray variability of quasars and blazars}
X-ray observations of many quasars and blazars reveal very high variability on various time scales - from flares on 
minutes to hours to secular changes over days and years. 
McHardy (2010) reports analysis of X-ray data of 3C273, one of the most powerful quasars with a jet, collected from a 
large number of satellites over several years. The PSD of 3C273 seems exactly like that of a 
soft state Seyfert galaxy implying that the source of the variations $-$ ``note, not the source 
of the X-rays'' $-$ lies outside the jet and is simply modulating the emission from the jet e.g., variations 
propagating inwards through the disc (McHardy 2010). The jet is likely to be seen just as an extension of the corona which 
dominates the emission in Seyferts (McHardy 2010).

A rich and complex behavior is seen during X-ray flares from blazars. 
Observations of Mrk 421 reported by Takahashi et al. (1996), which looked at the time structure of flares 
in various energy bands, were the first to reveal that the soft X-rays lagged behind the hard X-rays. 
Synchrotron cooling where the lower energy electrons cool more slowly than the 
higher energy electrons, was invoked to explain why the soft X-rays lagged the harder X-rays.
An opposite trend was also seen in subsequent observations (Takahashi et al. 2000). 
This was attributed to particle acceleration since synchrotron cooling could not explain this.
In this scenario, the acceleration of fresh particles to higher energies leads to these particles radiating 
first in soft X-rays and later in hard X-rays. 
Later on, longer and continuous observations by Brinkmann et al. (2005) revealed the presence of 
both kinds of lags in Mrk 421, with timescales of $\sim$10$^3$s. This behaviour was
confirmed by Tramacere et al. (2009).  A lag of hard X-rays has also been reported from
another blazar, 1ES 1218+304, during a giant flare (Sato et al. 2008).
Moraitis \& Mastichiadis (2011) have used a two-zone acceleration model to explain
the flaring behaviour of 1ES 1218+304. They assume that "the synchrotron losses dominate 
the energy loss mechanism and that the loss of electron energy by the inverse Compton scattering 
can be ignored".  Their model can produce "a variety of flaring patterns by varying the rate of injection
of the particles in the acceleration zone". According to Moraitis \& Mastichiadis their model may 
be "applicable to the sources where the energy density due to magnetic fields is more than due to photons".

\section {Cosmic X-Ray background and heavily absorbed AGN}
Discovered two years before the better-known microwave background
(Giacconi 1962) cosmic X-ray background (CXB) covers a very wide energy range, 
extending from $\sim$0.1 keV to several hundred keV.  In the very soft X-ray band  ($\leq$ 0.30 keV), 
more than 90\% of the CXB likely originates as thermal emission in a local bubble of hot, 
optically-thin ($\sim$10$^6$ K) gas in the solar neighbourhood (Singh et al. 1983). Between 2$-$10 keV,
a power-law with a photon-index $\Gamma$= 1.4 [N(E)=E$^{-\Gamma}$] was usually found to fit the data.
HEAO-1 provided a more precise measurement of the spectrum of the CXB over the range 3$-$50 keV, 
that was better fitted by a diffuse 40 keV thermal bremsstrahlung model than a power-law (Marshall et al. 1980). 
It also found that at these energies the CXB is isotropic to within a few per cent on large scales, 
after account is taken of a weak Galactic component as well as the dipole radiation field due to the motion of our Galaxy. 
However, its origin in the diffuse hot intergalactic medium is no longer supported and recent observations 
with XMM-Newton and Chandra have shown that the integrated emission from a wide distribution of AGN 
at various redshifts and with different obscuring column densities can account for the bulk of the CXB. 
The highly perceptive observation made by Friedmann \& Byram with regard to CXB way back in 1967, 
has thus been proven to be correct for the energy range of 2$-$10 keV, even though uncertainty of 
$\sim$30\% remains on the exact normalisation of the X-ray background intensity due to cross-calibration 
difficulties between various satellite missions, at least 75\% of the CXB between 2$-$10 keV is now resolved
(Mushotzky 2000).  A significant percentage of CXB still unaccounted for, if one considers only 10$-$50 keV 
energy band (Worsley et al., 2004), and this could arise from truly diffuse emission or be due to distant 
and unresolved heavily obscured / Compton-thick AGN (Frontera et al. 2007) as their spectra can mimic the 
residual spectrum of the CXB.
Indeed a large numbers of so-called type-2 QSOs, or AGN with intrinsic luminosities similar to those of 
local population, but hidden behind large obscuring columns of log(NH)=22 or higher are being discovered
with X-ray satellites (Gandhi et al. 2004; Zakamska et al. 2004.) X-ray bright optically normal galaxies 
(XBONGs: see Comastri et al. 2002; Miolino et al. 2003) found in luminous, as well as faint, X-ray sources,  
possibly constituting $\sim$10\% of all X-ray selected AGN may be the other significant contributors to the hard X-ray
CXB.  Heavily absorbed AGN may outnumber their unobscured counterparts by a large factor, and it is 
likely that their numbers may have been underestimated because of the difficulty of detecting and identifying them.

\section {Future}
Multi wavelength studies of existing well known AGN, as well as distant and faint populations of AGN being 
discovered, are continuing to reveal a variety of activities in AGN, both in the spectral nature of these objects as well
as the variability characteristics across the wavebands from simultaneous studies.
This is making the overall picture more complex at the same time revealing a lot of information required to understand
not only their violent activity but also about their life-cycles. How do the SMBH in AGN grow? What is the symbiosis 
of growth between the SMBH and their hosts? Does their life-cycle follow a monotonic path of growth or does it evolve
through stages of quiet and resurgent periods? The expected launch of future X-ray missions like ASTRO-H and ASTROSAT, and the expected new generation of extremely large optical telescopes might have an important impact on understanding the physics of the multifrequency behaviour of AGN.

\end{document}